\documentclass[a4paper,12pt]{iopart}
\usepackage{amssymb}
\usepackage{latexsym}
\usepackage{a4}
\renewcommand{\i}{{\rm i}}
\renewcommand{\d}{\partial}

\renewcommand{\submitto}[1]{\vspace{28pt plus 10pt minus 18pt}
     \noindent{\small\rm {\it #1}\par}}

\begin{document}
\title{A critical analysis of the hydrino model}
\date{\today}
\author{A. Rathke\footnote{
ESA Advanced Concepts Team (DG-X), ESTEC Keplerlaan 1,
2201 AZ Noordwijk, The Netherlands
}}
\ead{andreas.rathke@esa.int}

\begin{abstract}
Recently, spectroscopic and calorimetric observations of hydrogen
plasmas and chemical reactions with them have been
interpreted as evidence for the existence of electronic states of
the hydrogen atom with a binding energy of more than 13.6\,eV.
The theoretical basis for
such states, that have been dubbed hydrinos, is investigated. 
We discuss both, the novel deterministic model of the
hydrogen atom, in which the existence of hydrinos was predicted, and
 standard quantum mechanics. Severe inconsistencies in the
deterministic model are pointed out 
and the incompatibility of hydrino states
with quantum mechanics is reviewed.
\end{abstract} 

\submitto{New Journal of Physics {\bf 7} (2005) 127}
\pacs{03.65.Ta,31.10.+z,32.30.-r,36.90.+f}
\maketitle

\section{Introduction}

Recently experimental results have been published in respectable
physics journals that have been interpreted in support of a new
model of the hydrogen atom \cite{NJP1,NJP2,JAP1,JAP2}. This model
predicts the existence of new orbital states for the electron of the
hydrogen atom with enhanced binding energy compared to the known
hydrogen ground state. These new states have been named hydrinos.
Applications of these alleged states have already been
considered. In particular the NASA Institute for Advanced Concepts has
funded a study to investigate new propulsion concepts based on
the transition of conventional hydrogen states to hydrino states
\cite{NIAC}.

Although the hydrino model has received considerable public attention,
the discussion of the underlying theory has mainly been restricted to
the sweeping statement that the hydrino model is in contradiction to
quantum mechanics and hence dubious (cf.\ e.\,g.\ \cite{nature}). This
lack of theoretical consideration is particularly unfortunate in view
of the wealth of experimental evidence that has been published in peer
reviewed journals in favour of the hydrino model
\cite{NJP1,NJP2,JAP1,JAP2,
exspec,
dark,
vibration,
Ar,
Rb,
RbII,
inversion,
balance,
balance2,
normallines,
normallines2,
KHI,
KHI2,
MH,
SiH,
mini}. In this
paper we attempt to fill this gap by giving a brief review of the
theory underlying the hydrino model. We investigate its internal
consistency and comment on the possibility that hydrino-like states
exist in standard quantum mechanics.

Hydrinos are alleged lower-energetic electronic states of the hydrogen
atom. These states are predicted within a new deterministic theory of
quantum mechanics called the ``grand unified theory of classical
quantum mechanics'' (CQM) \cite{book}. In this theory the sheath
electrons of an atom are orbiting the core at a fixed distance on a
so-called orbit sphere. For the well known electronic states of the
hydrogen atom the radius of the orbit sphere equals the radius of the
corresponding state in Bohr's model. For the new hydrino states, the
radius is $r=q \, a_H$, where $a_H$ is the Bohr-radius and $q$ is a
pure fraction.  The corresponding binding energies are given by $W_q =
W_1 /q^2$ where $W_1 = 13.6\,$eV is the energy of the standard
hydrogen ground state.  The standard ground state of the hydrogen atom
is assumed to be metastable and the new hydrino states are only
attainable by ``non-radiative'' transitions
\cite{book,superfluid}. These states are assumed to be reachable in
the collision of hydrogen atoms with a catalyst, which can make an
electronic transition of the same energy. In the collision the energy
is transfered from the hydrogen to the catalyst, which absorbs it by
an electronic transition to a more energetic state. Eventually the
catalyst will release the acquired energy by the emission of a photon
and return to its ground state. The lowest-energy hydrino state, the
real ground state of the hydrogen atom, is then determined by the
requirement that the orbital velocity of the sheath electron must not
exceed the speed of light. The use of the alleged hydrino states for
power systems relies on inducing the decay of hydrogen to a hydrino
state and using the energy released in this process.

The layout of this paper is the following. In section~\ref{cqm} we assess
the underlying theory and review its internal consistency.  In
section~\ref{standard} we briefly consider the possibility that states of
the hydrogen atom with enhanced binding energy could exist in standard
quantum mechanics.
We summarise our conclusions in section~\ref{conc}.

\section{Aspects of the so-called ``Grand unified theory of classical
  quantum mechanics'' \label{cqm}}

This section is devoted to a review of R.\, Mills' ``grand unified
theory of classical quantum mechanics'' (CQM), which he claims to be a
consistent, deterministic, and Lorentz invariant replacement of
standard quantum mechanics \cite{book}. The theory allegedly predicts
the existence of new lower-energy states of the hydrogen atom ---
hydrinos. Brief expositions of the theory can be found in
\cite{cqm,superfluid}.  In this theory, the equation of motion of a
charged elementary particle, in particular that of the electron, is
given by a scalar wave equation with a peculiar dispersion
relation. To obtain the bound states of the electron in the hydrogen
atom this equation of motion is augmented by a quantisation condition,
which is similar to that of Bohr. In the following we discuss the
solutions for the wave equation presented in \cite{cqm,superfluid},
which are subject to this quantisation condition. We pay particular
attention to the solutions associated with alleged new electronic
states of the hydrogen atom.

CQM assumes that the dynamics of the electron are described by a
classical wave equation for its charge-density function, $\rho(t,{\bf x})$,
\begin{equation}
\left(
\nabla^2 - \frac 1{v^2} \frac{\partial^2}{\partial t^2}
\right) \rho (t,{\bf x}) = 0 \, ,
\label{wave}
\end{equation}
where $v$ is the phase velocity of the wave.

Already this starting point is troublesome in view of the fact that
this wave equation is not Lorentz invariant for any other phase
velocity than the speed of light. Hence we find, in
contrast to the claims of \cite{cqm}, that the
theory can at best be the non-relativistic limit of a broader theory,
but more probably is inconsistent already from equation\ (1) of \cite{cqm}.

For the following we will put aside these concerns and focus on the part
of the theory which is essential to the existence of hydrinos. This
is the solution of the wave equation (\ref{wave}) for the
hydrogen atom. CQM postulates a particle--wave duality, 
in which the wavelength of the solution of the wave equation 
has to correspond to the classical circumference of 
the electron orbit as derived from the classical motion of the
electron in a Coulomb potential, i.\,e.\
\begin{equation}
2 \pi r_n = \lambda_n \, ,
\label{rn}
\end{equation}
where $r_n$ and $\lambda_n$ denote the radius of the electron orbit
and the wavelength of the electron, respectively, and $n$ labels the
allowed orbits. Furthermore the de\,Broglie relation
between the wavelength, $\lambda$, and momentum, $p$, of a particle is
assumed to be valid,
\begin{equation}
\lambda = h/p \, ,
\end{equation}
where $h$ denotes Planck's constant. Combining the two relations, one
obtains for the phase velocity of the electron in the $n$th orbit,
\begin{equation}
v_n = \frac{\hbar}{m_e r_n},
\label{vn}
\end{equation}
where $m_e$ is the mass of the electron and $\hbar \equiv h/2\pi$.

If you combine the relations in equations (\ref{rn})--(\ref{vn}) with the
classical circular motion of an electron in the Coulomb field of a
proton, the ground state of Bohr's model is the only solution. No
solutions exist for excited states of the hydrogen atom. In order to
obtain the whole set of radii of Bohr's model one would have to change
equation (\ref{rn}) to $2 \pi r_n = n \lambda_n$, where $n$ is a positive
integer. Disregarding this fact, R.~Mills claims that the solutions to
the wave equation for the electron of the hydrogen atom are given by
\begin{equation}
\rho(r,\theta,\phi,t)
= \frac e{4\pi r^2}
\delta (r - r_n) \, Y^0_0 (\theta,\phi)
\label{l0sol}
\end{equation}
for zero orbital angular momentum, and
\begin{equation}
\rho(r,\theta,\phi,t)
= \frac e{4\pi r^2}
\delta (r - r_n) \, \left[ Y^0_0 (\theta,\phi)
+ \Re \left(
Y^m_l (\theta,\phi) \left[ 1 + \exp (\i \omega_n t)
\right] \right) \right]
\label{lxsol}
\end{equation}
for non-zero orbital angular momentum, $l>0$, where $r_n$ is the
radius of the $n$th orbit in Bohr's model.  Here $\delta(x)$ denotes
Dirac's delta function, $\Re$ denotes taking the real part of the
following expression, $Y_m^l$ denote the spherical harmonics,
$r,\theta,\phi$ are the spherical coordinates in obvious notation and
$e$ is the electron charge. Assuming that the radii $r_n$ can be
obtained by some other procedure (which Mills does not specify), we
check if Eqs.\ (\ref{l0sol}) and (\ref{lxsol}) are at least solutions
of the wave equation with $v=v_n$.  In \cite{cqm} it is stated that
the ``solutions'', Eqs.\ (\ref{l0sol}) and (\ref{lxsol}), can be
obtained by a separation ansatz,
\begin{equation}
\rho(r,\theta,\phi,t) = f(r) \, A(\theta,\phi,t) 
= f(r) \, Y(\theta,\phi) \, k(t)\, .
\end{equation}
Using this ansatz we can transform the wave equation into
\begin{equation}
\frac 1{r^2\,f(r)} \frac{\d}{\d r} \left( r^2 \frac{\d}{\d r}\right) f(r) 
= - \frac 1{A(\theta,\phi,t)} \left(
\frac 1{r^2} \triangle_{\theta,\phi} + \frac 1{v_n^2} \frac{\d^2}{\d t^2}
\right) A(\theta,\phi,t) \, ,
\label{firstsep}
\end{equation}
where $\triangle_{\theta,\phi}$ denotes the angular part of the
Laplace operator. Furthermore \cite{cqm} states that the separation yields
\begin{equation}
\left(
\frac 1{r^2} \triangle_{\theta,\phi} + \frac 1{v_n^2} \frac{\d^2}{\d t^2}
\right) A(\theta,\phi,t) = 0 \, .
\label{angeq}
\end{equation}
We decide to accept this claim for the moment. Continuing the
separation into an angular part, $Y(\theta,\phi)$, and a time part,
$k(t)$ we find the equation for the function $k(t)$
\begin{equation}
\frac 1{\omega_n^2} \frac{d^2}{d t^2} k(t) = {\rm const} \times
k(t) \, ,
\label{time}
\end{equation}
where we have used $v_n = \omega_n r_n$, with $\omega_n \equiv 2 \pi
v_n/\lambda_n$. Equation (\ref{time}) has the solutions
\begin{equation}
k(t) = {\rm const} \, , {\rm \quad and \quad \quad} k(t) = e^{\pm \i a \omega t} \, .
\end{equation}
where $a$ is a constant.  With these solutions the differential
equation for the angular part becomes
\begin{equation}
\triangle_{\theta,\phi} \, Y(\theta,\phi) = 0
\end{equation}
for the time independent solution, $k=$const, and
\begin{equation}
(\triangle_{\theta,\phi} - a^2 )\, Y(\theta,\phi) = 0
\end{equation}
for the solution which is harmonic in time.  Hence for the
time-independent situation the angular function has the solution
$Y(\theta,\phi) = Y^0_0(\theta,\phi) = 1$. The time-dependent solution
$k(t)$ only allows solutions for $Y(\theta,\phi)$ in terms of
spherical harmonics if $a$ is imaginary and fulfils the equation $a =
\sqrt{- l(l+1)}$ with $l$ being a positive integer. Hence, we can recover the
angular part of equation (\ref{l0sol}) but not that of equation (\ref{lxsol}).
In conclusion, equation (\ref{lxsol}) is not a solution of the wave
equation (\ref{wave}).  In practical terms this inconsistency of
the theory means that the model cannot describe the electron motion in
a hydrogen atom with non-minimal angular momentum. Note that the
electron states with non-zero angular momentum are well-described in
standard quantum mechanics. Hence CQM lacks important features of
quantum mechanics and does not describe known physics properly.

More important for our considerations is the radial part of the wave
equation because it contends to admit solutions with an orbital
radius smaller than that of the ground state of Bohr's model.  From
equation (\ref{firstsep}) and equation (\ref{angeq}) we know that the radial
part of the wave equation is given by
\begin{equation}
 \frac{d}{d r} \left( r^2 \frac{d}{d r}\right) f(r) = 0 \, .
\label{radialwave}
\end{equation}
This is the well known Euler differential equation.
The general solution to this is (cf.\ e.\,g.\ \cite{Euler})
\begin{equation}
f(r) = c_1 + \frac{c_2}r 
\end{equation}
However \cite{cqm} and \cite{superfluid} give the solution 
\begin{equation}
f(r) = \frac 1r \delta (r-r_n) \, .
\label{Millsradial}
\end{equation}
Using the standard expression for the $n$th derivative of the Dirac
function, $\delta(x)$, (see e.\,g.\ \cite{Weisstein}),
\begin{equation}
\frac {d^n}{d x^n} \delta(x) = (-1)^n n!\, x^{-n} \delta(x) \, ,
\end{equation}
it is straightforward to check that equation (\ref{Millsradial}) is not a
solution of the radial part of equation
(\ref{radialwave}).  However equation
(\ref{Millsradial}) is also claimed to be the radial solution for
hydrinos \cite{superfluid}.
Since equation (\ref{Millsradial}) is not a solution of the radial part of
the wave equation (\ref{radialwave}) for any $r_n$ there is no
way of deriving the existence of hydrinos from the wave equation
(\ref{wave}).

Our analysis of the theory of \cite{cqm,superfluid} has demonstrated
that the theory is mathematically inconsistent in several points: the
quantisation condition of CQM allows only a solution for the ground
state of the hydrogen atom; the radial solutions for the charge
density function of the electron, as well as the angular solutions
with non-zero angular momentum, differ from those given in the
literature on CQM \cite{cqm,superfluid}. To uncover the latter problem
we did not resort to any physics argument but instead directly checked
the alleged solution of the underlying equations of motion. Hence
there is no way to cure the flaws of the theory by adding physical
assumptions. CQM is obviously inconsistent, and in particular does
not contain solutions that predict the existence of hydrinos. Hence we
can omit a further discussion of CQM and, in particular, will not
discuss the description of ``non-radiative'' electronic transitions.

\section{Hydrinos in standard quantum mechanics\label{standard}}

Having found that CQM does not predict the existence of hydrinos (and
is furthermore inconsistent) it is worth considering if standard
quantum mechanics would allow for the existence of new electronic states
of the hydrogen atom with enhanced binding energy. 

We start with a discussion at the level of the Schr\"odinger equation.
In \cite{superfluid} it was mentioned that the Schr\"odinger equation
has solutions with main quantum number $n<1$. If such states were
allowed by standard quantum mechanics then also the existence of
hydrinos would also be possible in the standard theory. However, while
solutions of the Schr\"odinger equation with $n<1$ indeed exist, they
are not square integrable. This does not only violate one of the
axioms of quantum mechanics, but in practical terms prohibits that
these solutions can in any way describe the probability density of a
particle. Thus solutions with $n<1$ are meaningless in standard
quantum theory and the existence of hydrinos as solution of the
Schr\"odinger equation for a classical Coulomb potential is excluded.

The stability of the hydrogen atom in general is a long discussed
topic. For the isolated non-relativistic hydrogen atom, stability has
been proven, with a maximal binding energy of approximately 20\,eV
\cite{LiebH}.  This bound prohibits the existence of states with the
high binding energies attributed to hydrino states. The stability of
the hydrogen atom under the influence of external fields is not easily
proven in quantum field theory and the upper bound on the binding
energy is difficult to determine (see \cite{Lieb} for a review on this
problem).  Hence a state of the hydrogen atom that is lower-energetic
than the ground state cannot be ruled out completely under some exotic
conditions at our current level of understanding.  Such conditions are
however not likely to be fulfilled in the relatively low-energy, low
electromagnetic field environment of the plasmas studied by Mills et
al. (cf.\
\cite{NJP1,NJP2,JAP1,JAP2,exspec,balance,balance2,normallines}).  Of
course, there is no theoretical indication that the binding energy of
a putative new state of the hydrogen atom should be a multiple of that
of the ground state of the free hydrogen atom.  Note also that a
transition to a new state induced by strong external fields cannot be
a non-radiative one, which is in contrast to the interpretation of the
experimental data by Mills et al. \cite{NJP1,NJP2,JAP1,JAP2, exspec,
dark, vibration, Ar, Rb, RbII, inversion, balance, balance2,
normallines, normallines2, KHI, KHI2, MH, SiH, mini}. Hence, whereas
the stability of the ground state of the hydrogen atom is not yet
proven for all environmental conditions, the hydrinos have alleged
properties that make it impossible that their existence can be
encompassed by standard quantum mechanics.

\section{Conclusion \label{conc}}

In this paper we have considered the theoretical foundations of the
hydrino hy\-po\-thesis, both within the theoretical framework of CQM,
in which hydrinos were originally suggested, and within standard quantum
mechanics.  We found that CQM is inconsistent and has several serious
deficiencies. Amongst these are the failure to reproduce the energy
levels of the excited states of the hydrogen atom, and the absence of
Lorentz invariance. Most importantly, we found that CQM does not
predict the existence of hydrino states!  Also, standard quantum
mechanics cannot encompass hydrino states, with the properties
currently attributed to them.  Hence there remains no theoretical
support of the hydrino hypothesis. This strongly suggests that the
experimental evidence put forward in favour of the existence of
hydrinos should be reconsidered for interpretation in terms of
conventional physics. This reconsideration of the experimental data is
beyond the scope of the current paper. Also, to understand properly
the experimental results presented by Mills et al., it would be
helpful if these were independently reproduced by
some other experimental groups.

\end{document}